\documentclass[prl,twocolumn,showpacs,preprintnumbers,amsmath,amssymb,floatfix]{revtex4}

\usepackage{graphicx}
\usepackage{epstopdf}
\usepackage{dcolumn}
\usepackage{bm}

\begin{document}

\title{How \textit{Xenopus laevis} replicates DNA reliably even though its origins of replication are located and initiated stochastically}

\author{John Bechhoefer} 
\email[email: ]{johnb@sfu.ca}
\author{Brandon Marshall}
\affiliation{Department of Physics, Simon Fraser University, Burnaby, B.C., V5A 1S6, Canada}

\date{\today}

\begin{abstract}
DNA replication in \textit{Xenopus laevis} is extremely reliable, failing to complete before cell division no more than once in 10,000 times; yet replication origins sites are located and initiated stochastically.  Using a model based on 1d theories of nucleation and growth and using concepts from extreme-value statistics, we derive the distribution of replication times given a particular initiation function.  We show that the experimentally observed initiation strategy for \textit{Xenopus laevis} meets the reliability constraint and is close to the one that requires the fewest resources of a cell.
\end{abstract}
\maketitle

DNA replication is one of the defining processes of living systems, and evolution has accordingly selected for highly reliable replication mechanisms.  The South African clawed frog \textit{Xenopus laevis} is an organism often used to study replication in eukaryotes \cite{blow01a}.  The replication of its embryonic cells is particularly interesting, as it corresponds to a ``stochastic limit," where the placement and initiation of the sites where DNA replication begins (``replication origins") show significant stochasticity \cite{hyrien93}.   As with humans, the \textit{Xenopus} genome contains approximately three billion bases \cite{blow01b}.  Just after fertilization, cells divide for twelve generations with an abbreviated cell cycle that is as short as 25 min. (at 20 $^\circ$C).  The cell cycle is divided into an ``S phase" of about 20 min., when DNA is replicated, and a mitosis phase of about 5 min., when chromosomes separate and the cell divides \cite{hyrien93}.  In order to replicate so many bases in so little time, the cell initiates DNA replication at many [$\sim\mathcal{O}(10^5)$] origins.  For these embryonic cells, in contrast to the situation for fully developed somatic cells, there is no sequence dependence to the location of replication origins \cite{hyrien93}.  In addition, each origin initiates stochastically, with no pre-determined time of initiation.  The stochasticity in the location and initiation of replication origins leads to a potential difficulty:  the typical time for replication is about 20 min., but the maximum allowable time is only 25 min.  In particular, embryonic cells lack the efficient checkpoint mechanisms  \cite{hyrien03} that somatic cells have to pause the cell cycle to allow for unusually slow replication.  The cell must replicate by the time it divides, or die.  But empirically, such a ``mitotic catastrophe" \cite{prokhorova03} is rare, $\lesssim 10^4$ replications \cite{hensey98}.  How can one reconcile the variations in S-phase duration due to the stochastic placement and initiation of origins with the high reliability of replication?

In the biological literature the above is known as the ``random-completion problem" \cite{blow01b} and has been an unsettled question for over twenty years \cite{laskey85,hyrien03,rhind06}.  In its simplest form, randomly placed origins imply an exponential distribution of origin separations and, hence, a small number of very large gaps that take a long time to replicate.  Two approaches to a solution have been advanced.  The first notes evidence that the spacing of origins is not completely random and that any regularity in the spacing of origins will tend to suppress large gaps \cite{blow01b,jun04}.  However, in isolation, such a scenario is fragile:  if a single origin fails to initiate, it will create a much larger gap than exists usually.  The second approach draws on a recent experimental result that origins initiate throughout S phase and, indeed, that the rate of initiation of origins, $I(t)$ (initiations per time per length of unreplicated genome), increases significantly as S phase proceeds \cite{lucas00, herrick00,herrick02}.  Intuitively, initiating origins throughout S phase allows the cell to ``fill in gaps" and avoid unusually long delays.

In this Letter, we first calculate, following theories of nucleation and growth in one dimension \cite{sekimoto,ben-naim96}, the distribution of replication times $\rho_{rep}(t)$ given an initiation function $I(t)$ and a constant ``fork velocity" $v$ describing the symmetric growth of replication domains.  We find that an increasing $I(t)$ can insure replication at the required level of reliability, even in the worst case of completely random origin spacing.  We then show that the specific $I(t)$ observed in \textit{in vitro} experiments is close to an optimal $I(t)$ that minimizes the amount of cellular replication machinery (polymerases, helicases, etc.) that a cell is required to supply.

\begin{figure}[ht]
	\centering
	\includegraphics[width=2.5in]{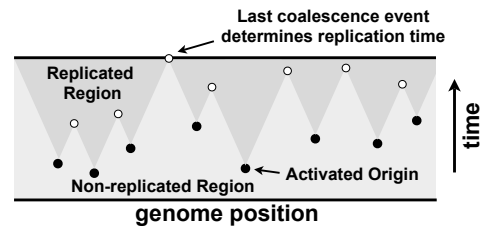} 
	\caption{Schematic of DNA replication model.  Space-time diagram showing multiple origins (filled circles), each expanding symmetrically at constant velocity.  Domains coalesce when they meet (open circles).}
	\label{fig:schematic}
\end{figure}

Our derivation of $\rho_{rep}$ uses a model inspired by the Kolmogorov-Johnson-Mehl-Avrami theory of crystallization kinetics \cite{christian02}, which is a stochastic model with three elements: nucleation (initiation) of ordered (replicated) domains; symmetric growth of these domains; and coalescence of domains that grow into each other.  (See Fig.~\ref{fig:schematic}.)  Using such a model, we showed that the fraction $f$ of DNA replicated on an infinite domain at a time $t$ after the start of S phase is given by
\begin{equation}
	f(t) = 1 - e^{-2vh(t)}
\label{eq:infinite-genome}
\end{equation}
where $h(t) = \int_0^t g(t') dt'$ and $g(t) = \int_0^t I(t') dt'$ and $I(t)$ is the initiation function ($\ge 0$) \cite{jun05a}.  Here, $v$ is the fork velocity, and $f(t)$ typically has a sigmoidal shape.  Equation~\ref{eq:infinite-genome} predicts that it will take infinite time to replicate all the DNA ($f=1$); but obviously, the replication time should be finite on a finite-length genome.  Because the location and time of initiation of origins is stochastic, the time to finish replication will also be a stochastic process.

In order to calculate the distribution of replication times $\rho_{rep}(t)$, we first note that, except for edge effects, there is a one-to-one mapping from replication origins to coalescences of replication domains.  (See Fig.~\ref{fig:schematic}.)  Because the evolution of domains is deterministic once the origin has initiated, one can derive the distribution of coalescence times, $\rho_c(t)$ from the initiation function $I(t)$.  In \cite{jun05a}, we derived the density of non-replicated domains (``holes") of size $x$ at time $t$ to be $n_h(x,t) = g^2(t) \exp[-g(t)x-2vh(t)]$.  Since a coalescence event is equivalent to a hole of zero size ($x=0$), we can write the normalized distribution $\rho_c(t)$ as
\begin{equation}
	\rho_c(t) = \frac{2vL}{N_o} g^2(t) e^{-2vh(t)} \; .
	\label{eq:coalescences}
\end{equation}
where $N_o$ is the total number of origins along a genome of length $L$ initiated throughout S phase.

As Fig.~\ref{fig:schematic} shows, the time to complete replication corresponds to the last coalescence event.  Since there are $N_o$ coalescences, the problem of determining the typical time of the last coalescence is equivalent to asking,  ``Drawing $N_o$ coalescences from a distribution $\rho_c(t)$, what is the largest time one expects to occur?"  Such questions are the subject of the field of extreme-value statistics \cite{gumbel58,kotz00}, where an analog to the central-limit theorem holds: given a parent distribution whose maximum value is unbounded and whose tail decays asymptotically at least as fast as an exponential (conditions satisfied here), the maximum value drawn in $N_o$ trials will, for $N_o$ large, tend to a Gumbel distribution, $\rho_G(\tau) = (1/\beta) \exp[-\tau-\exp(-\tau)]$, where the scaled time $\tau = (t- t^*) / \beta$, with $t^*$ the mode of the distribution and $\beta$ its width \cite{gumbel58}.  An elementary calculation \cite{gumbel58} shows that for Eq.~\ref{eq:coalescences}, the width $\beta$ is given by $2vg(t^*)$ and the
mode $t^*$ by
\begin{equation}
	F_c(t^*) = 1 - 1/N_o \; ,
	\label{eq:cumulative-tstar}
\end{equation}
where $F_c(t) = \int_0^t \rho_c(t') dt'$ is the cumulative probability distribution function (CDF) of the probability distribution function (PDF) $\rho_c(t)$.  From Eq.~\ref{eq:coalescences}, the CDF is, asymptotically for large $t$, given by
\begin{equation}
	F_c(t) = 1 - \frac{Lg(t)e^{-2vh(t)}}{N_o } \; .
	\label{eq:cumulative}
\end{equation}
Equation~\ref{eq:cumulative} is derived by integrating $\rho_c(t)$ by parts and dropping sub-dominant terms and, with Eq.~\ref{eq:cumulative-tstar},  leads to a transcendental equation for the magnitude of $I(t)$.

\begin{figure}[ht]
	\centering
	\includegraphics[width=3.0in]{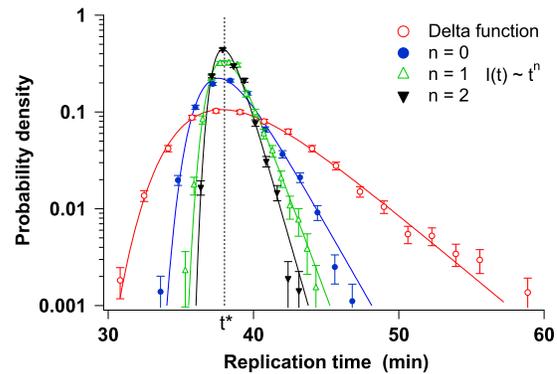} 
	\caption{Replication-time distribution function, fixing the mode to be $t^* = 38$ min.  Markers are results from Monte Carlo simulations (3000 trials per simulation); solid lines are fits to the Gumbel distribution.}
	\label{fig:t-star}
\end{figure}

In Fig.~\ref{fig:t-star}, we show the results of Monte-Carlo simulations of the replication-time distribution for various $I(t)$ functions.  In all cases, we adjusted the amplitude of $I(t)$ so that the mode of $\rho_{rep}(t)$ is at $t^* = 38$ min., which corresponds to the mode deduced from the $I(t)$ measured in the \textit{in vitro} experiments.  (For the \textit{in vivo} experiments, $t^* \sim 20$ min. \cite{hyrien03}.)  The solid lines are fits to a Gumbel distribution.  The parameters deduced (the $\beta$s) are consistent with the values predicted in the paragraph above.

The striking implication of Fig.~\ref{fig:t-star} is that one can vary the width of the replication-time distribution $\rho_c$ by choosing an initiation function $I(t)$ that increases throughout S phase.  Initiating all the origins at the beginning of S phase [$I(t) = I_\delta \delta(t)$] leads to the broadest possible distribution.  Exploring power-law initiation functions $I(t) = I_n t^n$ (with $I_n$ fixed by the $t^*$ constraint), we see that as one progresses from constant ($n=0$) to linear ($n=1$) to quadratic ($n=2$) initiation functions, the width of $\rho_c$ is progressively reduced.  The replication-time distribution can also be calculated using the experimental $I(t)$ \cite{herrick02} (not shown).  The experimental $I(t)$ is close to a quadratic curve and its distribution is indistinguishable from the $n=2$ case.

It would thus appear that the cell can have arbitrarily reliable replication (an arbitrarily narrow distribution $\rho_{rep}$) simply by arranging for its initiation curve to increase fast enough.  In fact, the situation is more subtle.  Even when all origins are initiated at the beginning of S phase, it is possible to replicate with arbitrary reliability simply by having enough origins.  While it is true that there will be a few unusually long gaps that will set the replication time, these gaps may be reduced arbitrarily if one starts with enough replication origins.  We thus propose an alternate way of viewing the random-completion problem:  Instead of fixing the number of origins and looking at the replication times for different strategies, we fix a time $t^{**}$ at which either a cell has finished replication or it dies.  Since evolution selects on the basis of mortality, the replication parameters ($I(t)$, $v$, the number of potential origins, etc.) should be a consequence of this selection, and not \textit{vice versa}.  Choosing $t^{**}$ to be the cell-cycle time (25) min. and allowing a failure rate of $10^{-4}$, we calculate, for various forms of $I(t)$, the replication parameters required to meet the reliability constraint.  (Our results depend only logarithmically on the failure rate.)

In order to compare with experiment, we must confront a further problem.  While the \textit{in vivo} replication time is estimated to be 20 min., the \textit{in vitro} experiments require nearly twice this time to replicate.  We must thus make additional assumptions to translate the \textit{in vitro} experimental results to the \textit{in vivo} situation.  In fact, we can do this with one simple assumption.  In earlier studies, it was assumed that the replication fork velocity $v$ is constant throughout S phase.  The original analysis of the \textit{in vitro Xenopus} data thus estimated an average fork velocity of 0.6 kb/min.  More recent work \cite{marheineke04} has shown that the fork velocity starts at 1.1 kb/min. at the beginning of S phase and then decreases monotonically to 0.3 kb/min. at the end of S phase.  We speculate that the longer time for the \textit{in vitro} S phase is caused by this reduction in fork velocity -- perhaps because some protein concentrations are not kept constant.  With this single modification -- $v = 1.1$ rather than 0.6 kb/min. -- we shall find results consistent with the \textit{in vivo} observations.

\begin{figure}[ht]
	\centering
	\includegraphics[width=3.0in]{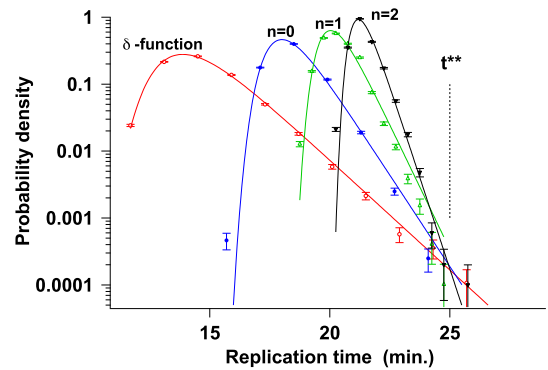} 
	\caption{Replication-times distribution function, fixing the mortality rate at $t^{**}$ = 25 min. to be $10^{-4}$.  (The area to the right of the dashed line of each probability distribution function is $10^{-4}$.)    Markers are results from Monte Carlo simulations (20000 trials per simulation); solid lines are fits to the Gumbel distribution.}
	\label{fig:t-dstar}
\end{figure}

In Fig.~\ref{fig:t-dstar}, we show results of simulations that constrain the replications to finish by $t^{**} = 25$ min., allowing a failure rate of $10^{-4}$.  We see that it is indeed possible to find amplitudes for $I(t)$ that satisfy the reliability constraint.

While it is always possible to choose an amplitude (e.g., $I_\delta$ or $I_n$) to satisfy the reliability constraint, each choice will have definite implications for the amount of cell resources that are required for its implementation.  One may then ask whether there is a ``best" strategy for initiating origins (while satisfying the reliability constraint).  If so, how close is the experimental $I(t)$ to the optimum?

To answer such questions, one must first define a measure for cell resources.  We have considered two possibilities among many that can be imagined:  the number of origins initiated throughout S phase and the maximum number of replication forks required.  The first choice would be relevant if the origin-initiation proteins were limited.  The second would be relevant if the number of polymerases (or other parts of the replication machinery) that needed to be active at one time limited the rate of replication \cite{rhind-private}.  We find qualitatively the same results in both cases \cite{marshall-unpub}.

Intuitively, there should be an optimum for the consumption of resources.  Within the fork-density scenario, initiating all origins at the beginning leads to a high initial fork density.  Holding off initiating until later in S phase helps by allowing the machinery of replication forks to be repeatedly reused.  If the cell waits too long to begin replication, then it is essentially shortening S phase, which requires many origins (and forks).  Thus, one expects an optimum.  We have explored this by calculating the maximum number of forks, $n_{max}$, required in several cases.  First, we calculated it for delta-function initiation ($n_{max} = I_{delta}$).  Next, we numerically calculate $n_{max}$ for the power-law case.  Finally, we use the calculus of variations to calculate the optimal $I(t)$, denoted $I_{opt}(t)$ that minimizes the maximum number of required forks, subject to the reliability constraint.   To calculate $I_{opt}$, we note that the number of replication forks is given by $n(t) = \dot{f}/v = 2g(t) \exp{-2vh(t)}$ \cite{jun05a}.  One can extract the maximum fork density using a technique familiar from control theory ($\mathcal{H}_\infty$ metric)  \cite{skogestad05}.  We thus write
\begin{equation}
	n_{max}[I(t)] = \lim_{p \to \infty}
	\left[ \int_0^\infty \left[ 2g(t) e^{-2vh(t)} \right]^p dt \right]^{1/p} \; .
	\label{eq:calc-variations}
\end{equation}
The associated Euler-Lagrange equation turns out to be independent of the exponent $p$.  We find
\begin{equation}
	\ddot{h}(t) = 2v \dot{h}^2(t) \; ,
	\label{eq:euler}
\end{equation}
where we recall that $\ddot{h}(t) = I(t)$ and $\dot{h}(t) = g(t)$.  Solving Eq.~\ref{eq:euler} subject to the boundary condition $h(0) = 0$ gives
\begin{equation}
	I_{opt}(t) = \frac{1}{2vt^*} \left[ \delta(t) + \frac{1}{t^*} \frac{1}{(1 - t/t^*)^2} \right] \;  .
	\label{eq:i-opt}
\end{equation}
Equation \ref{eq:i-opt} implies that the fork density $n = 1/vt^*$ is constant throughout S phase.  

\begin{figure}[ht]
	\centering
	\includegraphics[width=2.5in]{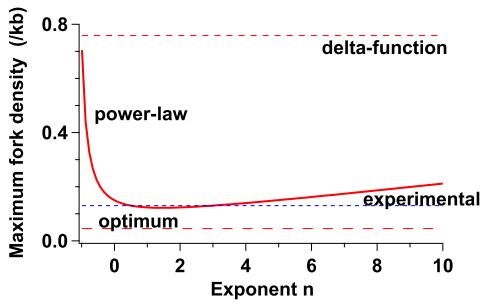} 
	\caption{Maximum required fork density, for different replication schemes.}
	\label{fig:fork-density}
\end{figure}

In Fig.~\ref{fig:fork-density}, we summarize the results of these investigations.  The dashed line at the top gives the fork density required to make the delta-function $I(t)$ meet the reliability constraint.  The solid curve represents the fork density required for power-law initiations.  As we anticipated, the curve has a minimum (between $n=1$ and 2).  The fine-dashed line, which lies close to the minimum value of the power-law case, is the experimental maximum fork density \cite{herrick02}.  Finally, the broad-dashed line gives the optimal fork density (1/$vt^*$).

Although the optimal fork density is lower than that observed, it clearly does not represent a physiologically possible case.  It is unrealistic to expect the perfect coordination implied by the delta function at the beginning of S phase.  More serious, at the end of S phase, Eq.~\ref{eq:i-opt} implies that the rate of initiation diverges, along with the total number of activated origins.  Still, we note that the qualitative shape of the curve shares the quadratically increasing form of the experimental result.  More generally, it would be surprising if the initiation program were identical to the optimum (even if one were to limit the space of functions to those that are physiologically achievable).  We note that the minimum is clearly broad:  there is little difference in required fork density between a linear  and a quadratic $I(t)$.  The main point is that there are some strategies -- most notably the initiation of all origins at the beginning of S phase -- that are clearly bad, and these differ from the observed $I(t)$.

In conclusion, we have calculated the distribution of replication times $\rho_{rep}$ for the stochastic limit of replication, where origins are placed randomly and initiate stochastically at a rate $I(t)$.  Choosing an $I(t)$ that increases with time narrows $\rho_{rep}$ and increases the reliability of replication.  Using the known mortality rates and length of the cell cycle, we gave a quantitative interpretation to the random-completion problem and showed that one can meet the reliability constraint using an arbitrary $I(t)$.  Different $I(t)$ functions demand different resources from the cell.  Measuring this resource use by the maximum required fork density, we show that the experimentally observed form of $I(t)$ is close to optimum.  In the future, it would be interesting to consider the effects of any regularity in origin spacing.  While we have shown that reliable replication may be achieved even in the worst case of random spacing of origins, there is evidence for some regularity.  It would also be interesting to measure the replication-time distribution directly.  While determining the time at which the last base (of three billion) replicates is unrealistic, one might be able to determine when a given fraction (e.g., 90 or 95\%) of origins have replicated.  It is straightforward to generalize the methods presented here to determine the distribution of times required to reach a given replication fraction.

We thank O.~Hyrien, N.~Rhind, R.~Harland, J.~Herrick, and S.~Jun for helpful discussions.   This work was supported by NSERC (Canada) and a visiting professorship at the Univ.~de Rennes, 1 (France).  JB thanks Z.~Gueroui for the invitation to Rennes.

\end{document}